\documentclass[pre,showpacs]{revtex4-1}

\usepackage[latin9]{inputenc}
\usepackage{amsmath}
\usepackage{amssymb}
\usepackage{graphicx}
\usepackage{geometry}                			
\usepackage{amssymb}
\usepackage{amsmath}
\usepackage{graphicx}
\usepackage{dsfont}

\usepackage{epsfig}\usepackage{color}\usepackage{booktabs}

\makeatother

\begin{document}

\title{The Tangled Nature Model of evolutionary dynamics reconsidered: \\
structural and dynamical effects of trait inheritance.}
\author{Christian  Walther  Andersen and Paolo Sibani}
\affiliation{FKF, University of Southern Denmark, DK5230 Odense M, Denmark}

\begin{abstract}
 Based on the stochastic 
dynamics of interacting  agents which reproduce, mutate and die,
the Tangled Nature Model  describes key emergent features of  
biological and  cultural ecosystems' evolution. 
While trait inheritance is not included in many applications, i.e.
the interactions of an agent and those of  its mutated offspring
are taken to be  uncorrelated,  in the 
family of TNM models introduced in this work correlations of varying strength 
are parameterised by a positive integer  $K$.
We first show that    the interactions generated by our rule 
are  nearly 
independent  of $K$. Consequently, the structural and dynamical effects of trait
inheritance can  be studied independently
of effects   related to  the form of the  interactions.
We then show that changing $K$ strengthens the core structure of the ecology, leads to
population abundance distributions better approximated by log-normal
probability densities  and 
increases the  probability that a  species extant at
time $t_{\rm w}$ also survives at  $t>t_{\rm w}$. Finally,
survival probabilities  of species are shown to decay as  powers 
of the ratio $t/t_{\rm w}$, a  so called pure aging behaviour 
usually seen in  glassy systems of  physical origin. 
We find a  quantitative dynamical effect 
of trait inheritance, namely  that increasing the value of $K$ 
numerically decreases  the decay exponent of 
the species survival probability.
\end{abstract}
\maketitle

\section{Introduction}
Models of biological~\cite{Drossel01}  and 
 social~\cite{Castellano09}  evolution
often  involve networks of   interacting agents
 whose  dynamics  is  interpreted in  biological or cultural
 terms~\cite{Rikvold03,Axelrod97,Javarone13}.  
Among these, the Tangled Nature Model is  a relatively recent~\cite{Christensen02,Hall02}
but already well-studied~\cite{Rikvold03,Anderson04,Anderson05,Sevim05,Lawson06,Murase10,Becker14,Nicholson15,Arthur16} 
 agent based stochastic description of  evolutionary dynamics which   features 
 punctuated equilibria~\cite{Gould77},  a key  dynamical property 
 of, among others,  macro-evolutionary systems. The latter
  go through a series of macroscopically different metastable states, with rapid and 
  dramatic changes, here called \emph{quakes}, leading from 
  one state to the next.
Numerical investigations of the TNM show other  observed 
 properties of ecosystems, such as the `area law'~\cite{Lawson06} and indicate that
 the model  is also applicable  to  human cultural and industrial ecosystems \cite{Nicholson15,Arthur16}.
 As  a simple stochastic model the TNM   cannot  precisely predict the 
  development of an ecosystem from a given  initial condition.
 It  can however predict   its emergent properties  in  a variety 
 of different situations and  it therefore clearly deserves  further development   and analysis.

Interestingly, depending on one's choice of how  interactions 
are generated,  the TNM can either quickly  reach a stationary state 
characterised e.g. by the power-spectrum of population or diversity
fluctuations~\cite{Rikvold03,Murase10}
or it can enter an aging regime~\cite{Anderson04,Jones10,Becker14}  
similar  to that of glassy systems~\cite{Sibani14}. Specifically~\cite{Becker14}, 
a distribution of  interactions with finite   support
leads to stationary behaviour,  while aging ensues  if ever larger positive interactions become accessible 
on ever longer time scales. This is the case presently considered, where 
as seen in~\cite{Anderson04,Anderson05,Jones10,Becker14},
 the dynamics is driven by a series of non-equilibrium events, our quakes,
   through a sequence of macroscopically different metastable states,
 often called QESS (Quasi Evolutionary Stable States).

The set of interactions linking   a TNM individual
to  others is key to  its  reproductive success and
arguably constitutes  its  most important property.
Yet, in many studies, e.g.~\cite{Anderson04,Jones10,Becker14},
 the interactions of an individual
and those of its   mutated off-spring are unrelated, a rather 
unrealistic feature corresponding  to a point mutation turning
a giraffe into an elephant.
To address this issue, 
Sevim and Rikvold~\cite{Sevim05}  start out with  an interaction matrix consisting of Gaussian deviates
with  variance identical
 to the uniform distribution  previously used~\cite{Rikvold03} by one of these authors.
The  matrix is then averaged   locally 
over neighborhoods  in  genome space to produce the desired correlations.
 These authors find a  
 stationary fluctuation dynamics,   resembling  that  of a model with uncorrelated 
 interactions. We note that their  approach requires the storage and manipulation of huge 
 sparse matrices. 
  Laird and Jensen~\cite{Laird06} introduce correlated interactions 
    by representing 
  individuals  as   16 dimensional vectors in phenotype space. Each element is an
  integer in  the set $\{ 0,1, \ldots99999 \}$ and contributes additively to the interaction
  of two species. This leads to a Gaussian interaction distribution
  where  a change of a single element  has   a minor effect on the total, as 
  desired. 
  The Gaussian  distribution of interactions used in both descriptions quickly makes the appearance of 
  `destabilizers'~\cite{Becker14}, e.g. mutants receiving very strong positive interactions from
  extant species, extremely rare, whence the ecology
  evolves 
   at a pace  considerably  slower than in Refs.\cite{Anderson04, Jones10, Becker14}. 
   Choosing one type of interaction distribution over
   another has structural and  dynamical effects which seem to have been
   overshadowed by issues of numerical convenience
   and, consequently,  have not received
   sufficient attention in the literature.
    
 Below, we introduce and analyze  a one-parameter 
family of TNM models 
where an increasing degree of correlation
between the interactions  of an agent and those of its mutated offspring
is obtained by increasing the value of 
a positive integer  parameter, $K$, 
where  $K=1$  corresponds 
the model version without trait inheritance used in  Refs.~\cite{Anderson04,Anderson05,Jones10,Becker14}.
The  interactions  between individuals of different species  
    arise  as products of two  Gaussian variables
    of zero average. For $K=1$  these are independent  and 
 the  probability density function (PDF)    of the interactions is  a  modified Bessel function
    of zeroth order. For $K>1$, the two Gaussian variables independence is no\
 longer   mathematically guaranteed.
 Nevertheless, the  interactions generated by our rule turn out to be   largely insensitive to the value of $K$,
 and  the dynamical effects of
trait inheritance  can therefore  be investigated  independently
of any  effects  imputable to a change of the interaction distribution itself.

The rest of the  paper is organized as follows:
after summarizing  the background and notation, we first explain how  the  interactions are generated,
and estimate, for different values of $K$,  their  distribution  and the correlation  
between the interactions of an agent and those 
of its  mutants.
We then describe how trait inheritance  affects the dynamics, 
first qualitatively at the level of the emergent  \emph{core species}  structure~\cite{Becker14}
and then, more quantitatively, in terms of the species abundance distribution and the time decay of 
a cohort of species  picked at different stages of the system evolution.

\section{Background and notation}
TNM agents are  binary strings 
 which can either  be interpreted   as genomes  or  as 
cultural features\cite{Nicholson15,Arthur16}, i.e. blueprints or strategies for action.
Reproduction is  asexual and error prone, and its  rate   depends on  the  `tangle'  of interactions
 connecting  the agents, with positive, or mutualistic, interactions leading  to a higher reproduction rate.
 Removals happen  at a  constant rate and independently of the interactions.
Since  extant agents draw resources from a shared and finite   pool, they  all
have an   indirect, global and  negative effect  on each  other's reproductive success.
For sufficiently large values of the coupling probability $\theta$, and irrespective of the
 degree of correlation, a typical  TNM ecology comprises a single  group of interacting species. 
 
 As described
 in Ref.~\cite{Becker14},  the  TNM ecology  can be sub-divided into  a small group
 of populous  \emph{core} species and a majority of  intermittently populated \emph{cloud} species.
 The core is an \emph{ordered} structure,  since core species are inevitably
 linked by mutualistic interactions dynamically selected from a symmetric
 distribution. The network structure  spontaneously emerges in a 
 process driven by  an overall  \emph{increase} of 
  configurational entropy~\cite{Becker14}, showing that 
  entropy and order generation are not necessarily antithetic, as often surmised.
  In the TNM the two  grow 
  simultaneously in  different parts of the system: core species carry the order and cloud species 
 carry  the entropy.
  We finally note that the mutualistic interactions  between TNM  core species do not have a direct interpretation
  in terms of trophic chains. 
  
For completeness, we now
briefly summarize the notation  used e.g.  in Refs.~\cite{Jones10,Anderson04,Becker14}.
A species is a group of agents with identical genomes,
and  agent `$a$' is queried with  probability equal to the relative size of 
its species. When queried, the agent  reproduces with
probability 
\begin{align}
p_{\rm off}(a)= \frac{1}{1 + e^{-H_a}}, \quad {\rm where} \quad H_a = \frac{C}{N(t)}\left(\sum_i J_{ai} N_i\right) - \mu N_i.
\label{eq:offspring_prob}
\end{align}
In the rightmost expression, from left to right,  $C$ is a scaling constant, the coupling  $J_{ai}$ represents the
influence of agent $i$ on agent $a$,
$N_i$ is the current population of species $i$ and $\mu$ is 
a constant expressing the carrying capacity of the environment. Note that $J_{ai} \neq J_{ia}$ and that 
$J_{ii}=0$, i.e. self-interactions are excluded.
Each bit (gene) in the genome of a newly created off-spring differs from the parental gene
with a constant probability $p_{\rm mut}$.
A last parameter, $\theta$, determines the probability that two species are connected 
by non-zero interactions. 
Time $t$ is measured in generations, whose  length or duration  equals  the
number $N(t)/p_{\rm kill}$ of Monte Carlo queries   needed on average to remove all individuals 
present at time $t$  when  $p_{\rm kill}$ is the removal probability.
The number of Monte Carlo queries within 
  a generation is thus calculated iteratively 
based on the population of the previous generation.

The parameters used in this work are $\theta = 0.25$, $\mu = 0.05$, $C = 50$, $p_{mut} = 0.01$ and $p_{kill} = 0.25$. 
Initially,  the ecology  consists of single species of $500$ identical individuals.
The above  initial condition is standard in the literature and has little bearing on 
the observed statistical properties on longer time scales. Invariably, the population of a single--and thus non-interacting-- 
species quickly reaches the level
dictated by balancing its death and reproduction rates and 
the evolution of the ecosystem first starts once mutations have created 
a group of interacting species.
In our case, the system is  given   a couple of generations to 
find a metastable configuration before data collection begins.

The basic procedure to generate the  interactions between species `$x$' and `$y$'  
from their  corresponding  binary genomes of  length $L$, $G(x)$ and $G(y)$,  relies on 
both their binary representation and their  equivalent
 integer representation.
 Right below  and until further notice,   the name of a species, e.g. $x$,
and that of its genome, i.e. $G(x)$, will for convenience  be identified in the notation.

Two   random arrays of length $2^L$, $A$ and $B$, are 
initially constructed and never changed during a  simulation. 
$A$ contains  independent standard Gaussian deviates of zero average,
and $B$  contains independent binary variables which
equal zero with probability $1-\theta$ and one with probability $\theta$.
The logical \emph{exclusive or}, $z=$XOR$(x,y)$ is a binary string of length $L$ 
calculated by performing a bitwise  XOR operation on the binary representation
of $x$ and $y$.
The three strings $x$,   $y $ and $z$ are  finally  read as integer indices to 
the arrays $A$ and $B$. With an eye to the next section, where the notation is slightly 
more involved, we denote the $x$'th element of the array $A$ by $A(x)$ instead of the more
usual $A_x$.
 
The  procedure to generate $J_{xy}$ is
now  as follows:
\begin{enumerate}
\item Calculate $z=$XOR$(x,y)$.
\item If ($B(z)=0$ or $x=y$),  set  $J_{xy}=0$.\\ If not,
\item
\begin{enumerate}
\item Read out the three Gaussian deviates\\  $z_1=A(x)$, $z_2=A(y)$ and $z_3=A(z)$.
\item Set $J_{xy}=z_1 z_3$ and $J_{yx}=z_2 z_3$.
\end{enumerate}
\end{enumerate}
Note that  non-zero  interactions  are asymmetrical,  while zero interactions
are symmetrical.
Importantly, all interactions  are known \emph{in potentia}
before the dynamics starts and   never need to   be stored in  their totality. 
Computationally this is advantageous, since only  the 
 interactions between extant species are needed for updating the dynamical state 
 of the system.
  These  interactions remain  at all times a tiny
fraction of the existing $2^{2L}$ possibilities, but move around and acquire  more positive
values~\cite{Anderson05,Jones10,Becker14} as the ecology evolves.

When using  the above scheme,
 changing a single bit in the  genome of an individual  leads to  a  completely
different set of interactions to other individuals and to the mentioned lack of trait inheritance.
 In the next section we illustrate how the scheme can be improved  to
allow  a mutant to  inherit
some  traits of its parent without destroying its computational convenience.

\section{Trait inheritance and correlated interactions}
In order to introduce trait inheritance, a gene of length $L$ is first  subdivided
into $K\leq L$ contiguous parts  of 
 integer size $L/K$, each part  indexed  by an integer $s$, $1\leq s \leq K$.
Secondly, we return to our full notation and  let   $G_s(x,0)$ denote   the  binary string whose s'th part is identical to
the corresponding part  of $G(x)$, and whose  other bits are  zero.
To calculate the interactions between $x$ and $y$ we repeat the procedure described
in the previous section, except that we now 
read out $3K$ standard Gaussian deviates,
$z_{1,s}=A(G_s(x,0))$, $z_{2,s}=A(G_s(y,0))$ and $z_{3,s}=A(G_s(z,0))$, and 
 define
\begin{equation}
J_{xy} = z_1 z_2; \; J_{yx} = z_1 z_3, \quad {\rm where} \quad z_l=\frac{1}{\surd K} \sum_{s=1}^K z_{l,s}\;; \quad 1\leq l \leq 3.
\end{equation}
We note that  all  three $z_l$'s are  standard Gaussian deviates 
of zero average and that changing one bit in, say, gene $x$ only affects one of 
the $K$ contributions to $z_1$. This  produces the desired correlations between 
the interactions of similar genes.
We did not implement an analogous procedure to correlate
whether an agent and its one-point mutant have similar  sets of zero
interactions. 
Finally, since each part of the genome corresponds to $2^{L/K}$ integers,
only $K 2^{L/K}$  elements of   the array $A$ are utilized
out of the $2^L$ available.
For   $K$ near $L$ this introduces  undesired statistical  correlations between the $z_{l,s}$ values  
generated by the algorithm.

\begin{figure}
	\centering
	\includegraphics[width=0.6\textwidth]{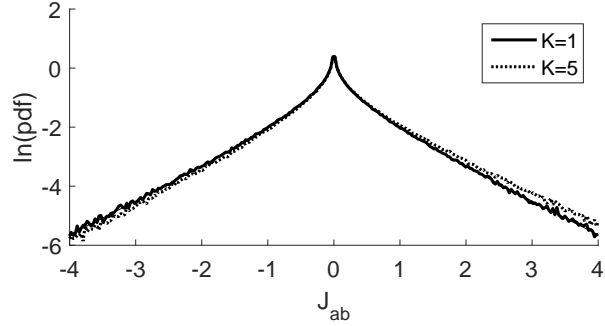}
	\caption{The PDF of the  interaction strengths for two different values of $K$.
	The theoretical PDF of the product of two standard Gaussian deviates with zero mean is also plotted but is indistinguishable 
	from the the $K = 1$ curve at the resolution level of the plot.}
	\label{fig:int_dist}
\end{figure}

The statistical properties of the coupling distribution are 
summarized in
 figure \ref{fig:int_dist}, which displays 
 an estimate of the PDF 
 of the interaction strengths  generated by
the rule just described for $L=20$ and for two values of $K$:
 $K=1$, corresponding to the uncorrelated model (full line) and  $K=5$. 
  For each $K$ value, our estimate of the interaction strength PDF 
  was  obtained by generating  $10^4$ arrays $A$ and $B$. For each of these couples of arrays, $10^4$ pairs
  of species  were generated as random binary strings of length $L$. The interactions between each pair  were then calculated as
  explained above. 
 The statistics is  thus obtained
 from  $2\times10^8$  non-zero interactions between random species. Hence it does  not include any dynamical 
 selection effects. The theoretical PDF of the product
 of two standard Gaussian deviates with zero average is also plotted.
 It  does however overlap completely   with  our empirical  $K=1$ curve and cannot be distinguished from it
  at the resolution level of the plot. 
 
As anticipated, the three PDFs have very similar shapes. The PDF for $K=5$ has  a slight positive bias,
at  very  high values of its argument. This is an effect caused, as mentioned,
 by only using  $K 2^{L/K}$ elements of the normally distributed array $A$:
 the probability of picking two random numbers with the same sign increases, and when these are multiplied together,
 the result is always positive.

\begin{figure}
	\centering
	\includegraphics[width=0.8\textwidth]{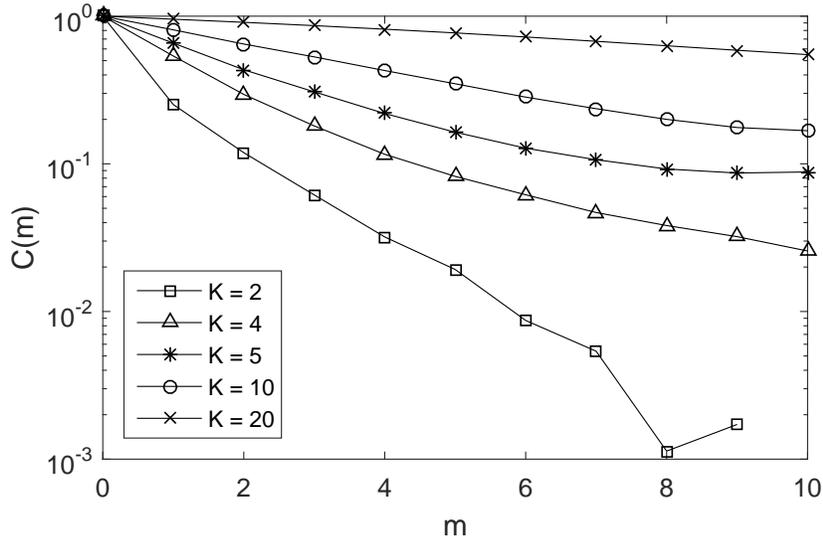}
	\caption{The normalised correlation function for species pair interactions  is plotted on a log scale as a function of 
	the number $m$ of point mutations which one of the two species undergoes. }
	\label{fig:corr}
\end{figure}

To describe how the interactions between two agents  $a$ and $b$ change as 
 $b$ undergoes mutations,  we first pick $10^4$ pairs $a,b$ of species and
  let $J_{ab}(m), \; m=0,1,2 \ldots$ be the interaction
between $a$ and an $m$ times mutated  $b$. 
We then define the correlation function 
$C(m) = \langle J_{ab}(0) J_{ab}(m) \rangle_{\{a,b\}}$, where the brackets 
indicate an average over the
 different 
$a$'s  and, 
for each $a$, over the $ L\choose m$ possible ways to introduce $m$ point
mutations in  $b$. For  $K=1$ the correlation function normalized to $C(0)=1$  is 
by construction a Kroneker delta, 
i.e. $C(m) = \delta_{m0}$. 
In fig.~\ref{fig:corr} similarly normalized correlation functions are  plotted  
on a logarithmic scale as  functions of $m$ 
for several $K$ values above $1$. 
The  functions  are  seen to decay with $m$  in a nearly
 exponential fashion and  the decay rate is seen to decrease with 
 increasing values of $K$.
 
In conclusion, our algorithm ensures that a mutant inherits the interactions of its
parent. The typical amount of  change induced by a mutation
decreases with increasing $K$.

\section{Dynamical effects}
Earlier studies \cite{Sevim05,Laird06} suggest that  correlated interactions
 do not qualitatively change  the dynamical behavior of the TNM. In this section we reexamine the question
 for the  model versions considered, focusing on structural features and  aging dynamics. 
 Our first analysis concerns the structure of the \emph{core species}~\cite{Becker14}
 and the log-normal distribution of  species abundance.
 We then proceed to analyze,  for different values of $K$,  the time dependence of 
 the survival probability of species which are extant at different ages. 
 \subsection{Core structure}
 \begin{figure}
$
 \begin{array}{lr}
 	\includegraphics[width=0.5\textwidth]{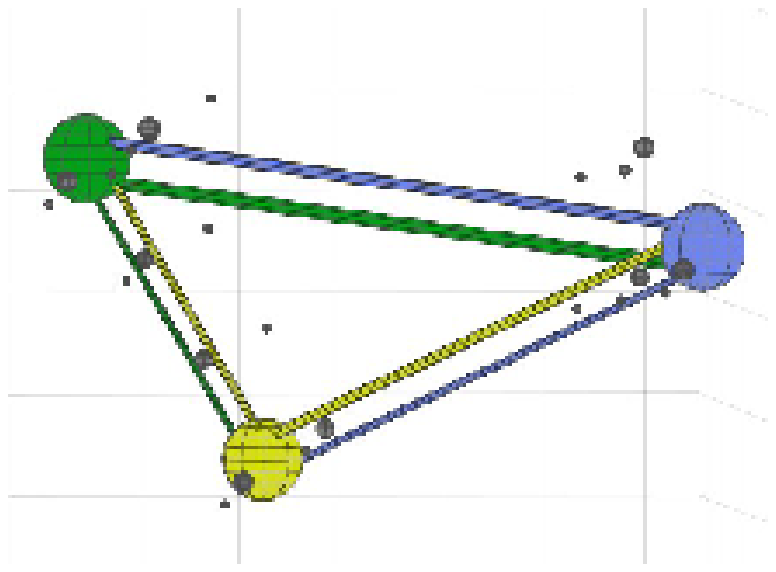} & \includegraphics[width=0.5\textwidth]{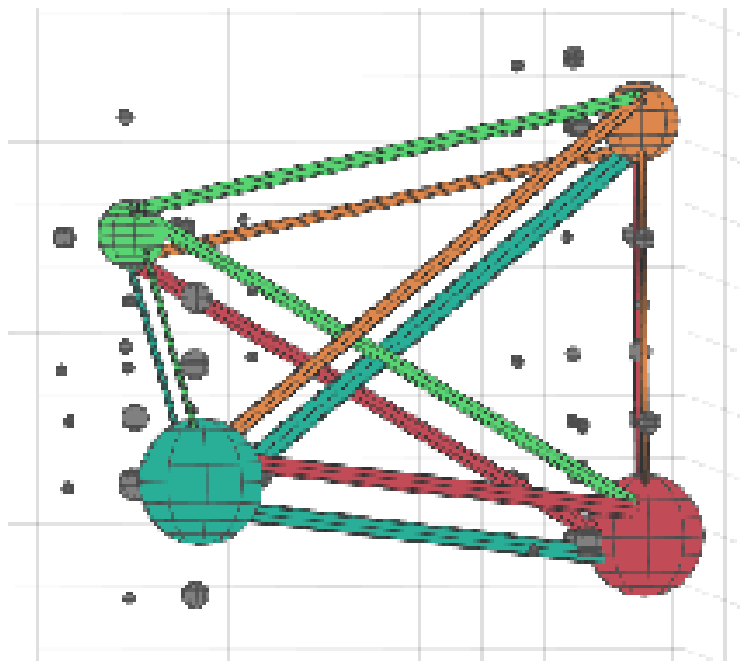}
\end{array}
$
\caption{(Color on line) A 3-dimensional  rendering  of a single  ecology after 10,000 generations.
	 Left, $K=1$ and right $K=5$.
	 Species are represented by spheres of volume proportional to their population.
	 The mutualistic interactions connecting core species  are  represented by
	 lines  with  the same color as the 
	 species they affect and with   a thickness proportional to their strength.
	 Interactions linking  cloud species are omitted for clarity.}
	\label{fig:corecloud}
\end{figure}

Core species make up the bulk of the population 
 since the mutualistic interactions which link them  together
  endow them with  high reproduction rates.
Most    species belong however to the cloud and 
are intermittently populated, mainly through an influx of mutants from
 nearby core species. All together, cloud species  only comprise 
 a minor fraction of the population. 
  Following~\cite{Becker14}, a practical criterion which can be
 used  to distinguish  cloud from core species on the fly is that
 a core species is   larger than
 5\% of the most populous species.

Figure~\ref{fig:corecloud} shows a three dimensional rendering of a single  TNM ecology 
evolved for $10^4$ generations starting from a single species with $500$ individuals.
The mapping   is  obtained using Principal Component Analysis, see e.g.~\cite{Nicholson15}
for further  details,  and represents species by spheres of volume proportional to
their  population.
Distances  reflect the Hamming distances 
between the corresponding species of the ecology. For $K=5$ the ecology appears more 
diverse than in the $K=1$ case where interactions are uncorrelated.
This seems a natural consequence of the fact that, in the former case,  a mutant species  inherits to some degree  the good
connections of its  parent core species and can more easily  establish itself as a new core species. This seems  to be  the case
for the two nearby core species drawn  near  the lower edge of the right panel.
Besides  one extra core species, the  $K=5$ version features many more cloud species
and a larger  total population.
\begin{figure}
	\centering
	\includegraphics[width=0.8\textwidth]{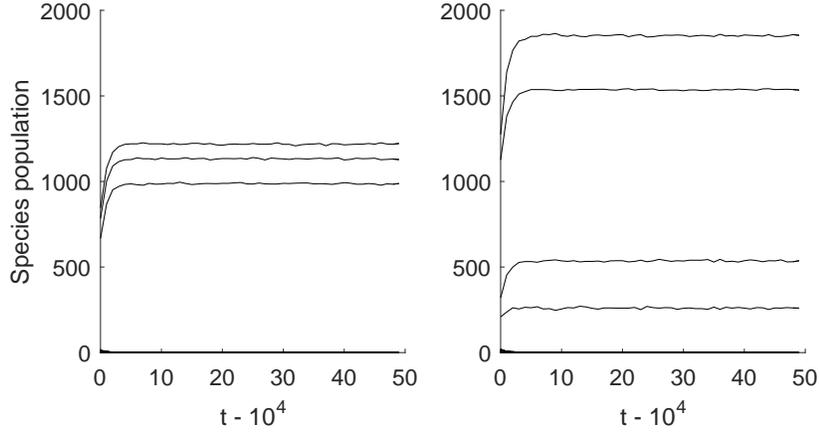}
	\caption{Each line represents the population of  a different species in the frozen ecology obtained by 
	instantanuosly setting the mutation rate
	to zero after $10^4$ generations.  The left and right panels
	show uncorrelated and correlated interactions, respectively.}
	\label{fig:frozen}
\end{figure}
\begin{figure}
	\includegraphics[width=0.4\textwidth]{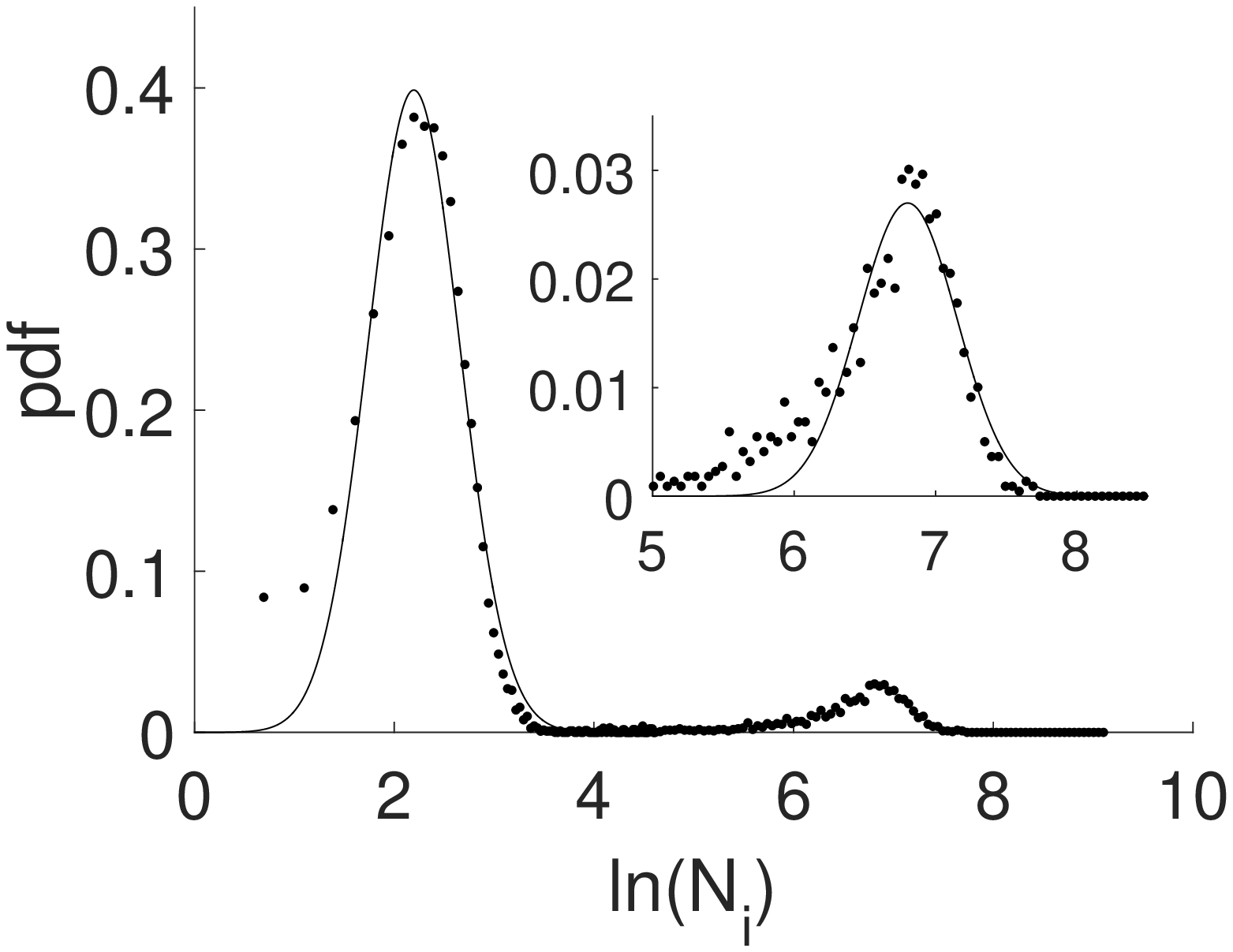}
	\includegraphics[width=0.4\textwidth]{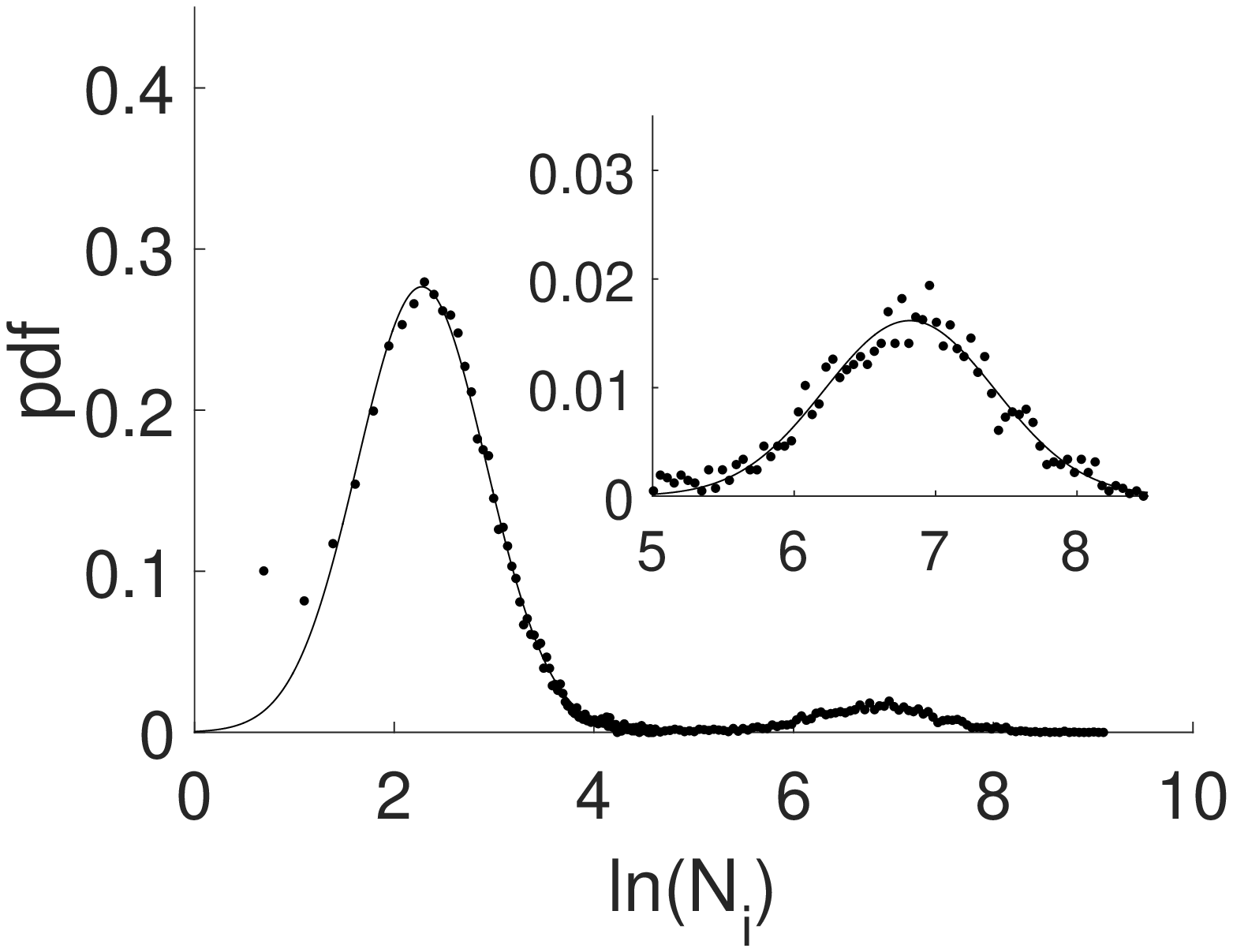}
	\caption{The population abundance distributions vs. the logarithm of the species population for  $K=1$ (left) and $K=5$ (right), respectively. Lines are gaussian fits. Notice the clear separation on the horizontal scale of the two  maxima of the distribution. Insets zoom in on  core species data at   large populations.}
	\label{fig:pop_dist}
\end{figure}
To show in more detail the differences in population and diversity induced by changing  $K$,
simulations were continued for a few generations  after forcing the  ecology to `freeze' at $t=10^4$,
by instantanuosly setting the mutation rate $\mu_{\rm mut}$ to zero. 
The immediate effect of removing mutations, seen in both panels of 
Fig.~\ref{fig:frozen}, is that core species    grow in size, due to the fact that 
all cloud species, no longer replenished by
mutants, die out. 
The long term effect is to force the dynamics into 
 a stationary state by 
 preventing   the generation of destabilising mutants.
Figure~\ref{fig:frozen} depicts  core species populations in such state  for $K=1$ and $K=5$ as a function
of time and shows that turning on the correlations increases both the diversity and the total population
of the ecology.
\subsection{Population abundance distribution}
Figure~\ref{fig:pop_dist} illustrates how the population is distributed across
different species.
The abscissa is the logarithm of the species population size, and the ordinate is our  estimate of the 
corresponding probability distribution function, obtained as the frequency with which a species of given size
appears in the population.
The data were obtained from an ensemble of $100$ independent runs each lasting for $10^4$ generations.
Comparing the left and right panels, we note that, irrespective of $K$,  the abundance distribution is  bimodal,
with two widely  separated maxima corresponding to values of the abscissa near $2$ and $7$.
The separation supports the distinction between 
cloud and core species.
The insets show that, for $K=5$, core species are distributed in a
nearly  log-normal fashion while,
for $K=1$, a  log-normal distribution still provides a reasonable, but  far less accurate, description.
We also see that the distribution of  the most populous  half of the cloud species is 
also  approximately log-normal, while  very sparsely populated
species fall outside the description. The nearly log-normal distribution of cloud species is expected,
as these species are populated by an influx of mutants from `parent'  core species
 with the size of the influx mirroring the size of the parent species.

Anderson and Jensen~\cite{Anderson05} already pointed out the relevance for  
the TNM of log-normal distributions,  which are known to describe many natural systems,
including   abundance distributions in several real ecologies~\cite{Limpert01}. 
Our results concur with  theirs  and  furthermore show \emph{i})  that the distinction between core and
cloud species emerges naturally from a statistical
analysis and \emph{ii}) that 
 introducing correlations improves the quality of the  log-normal
description of  the species abundance distribution.
\section{Species persistence curves}
Since, by definition, TNM agents die at a constant rate,  they  have a finite
expected life-time. In contrast,  and reflecting the complexity of the dynamics, species die  at a slow and decelerating rate, 
and their   life-times do  not possess finite averages. 
Studying species persistence curves throws  light on important aspect of the 
TNM dynamics: first, these curves concisely describe the aging dynamics of the model
and, second, they change in a measurable and systematic way
when inherited traits are introduced.

To explore these issues  we  define a cohort as
the set of  species   extant at time $t_w$. We then  measure the persistence  $P(t_w,t)$ as the 
fraction of the cohort which is still extant at time $t>t_w$.  The persistence provides an estimate
of the probability that a species extant at time $t_w$ still is  extant at later times and the distinction between the
 two quantities is glossed over in the following.
The life-time probability density function 
 of a species extant at 
$t_w$  is then
\begin{equation}
S(t_w,\tau)= - \frac{d}{d\tau} P(t_w,t_w+\tau) \quad 0\le \tau <\infty.
\label{survival}
\end{equation}
\begin{figure}
	\includegraphics[width=0.8\textwidth]{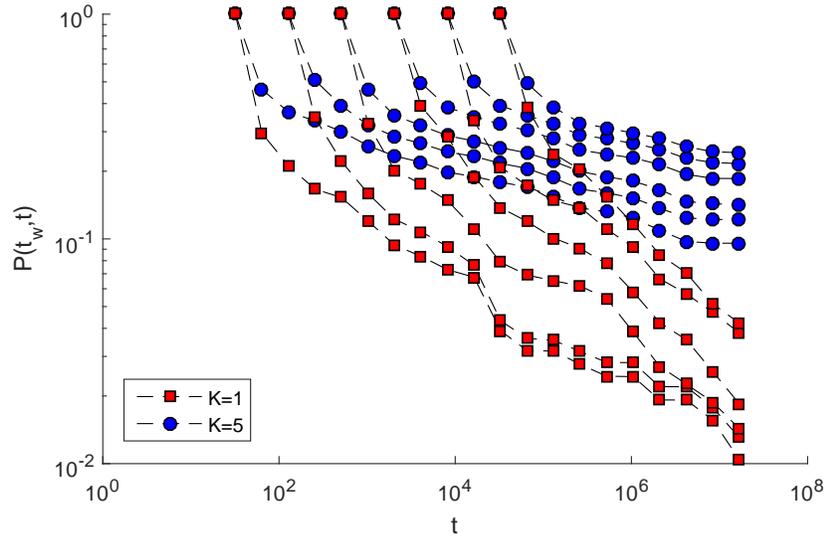}
	\caption{(Color on line) Species persistence  versus time.  A new   cohort of extant
	species  is registered for each $t = 2^k, k =1,2,3,  \dots$, and 
	the corresponding fraction surviving at later times is plotted on a logarithmic scale. 
	Not all cohorts are shown for graphical reasons. $K=1$ data
	are shown by squares, and $K=5$ data by circles.
	Lines are only guides to the eye.}
	\label{fig:survival-rate}
\end{figure}
To measure  persistence of species we  ran 51 independent simulations in the $K = 1$ case and 68 in the $K = 5$ case, all
 lasting  $2^{24}>10^7$ generations, which is a very long time by most criteria. 
 Extant species are registered at each $t = 2^k, k=4,5 \ldots 24$,
 with the delay on the first cohort introduced to ensure that the ecology has properly stabilised.
 We thus end up with $21$ cohorts, one for each $k$, of which  the last one is discarded
 as it only has one data point. 
Each data set in figure~\ref{fig:survival-rate} shows   for $t >t_w$ the persistence  of the cohort of species extant at time $t_w$.
The dip for $t \approx t_w$ stems from  the fast disappearance of cloud species. At later times the curves 
all tend to approach  straight lines on our  log-log plot, which is  indicative of a power-law dependence.
 \begin{figure}
	\includegraphics[width=0.8\textwidth]{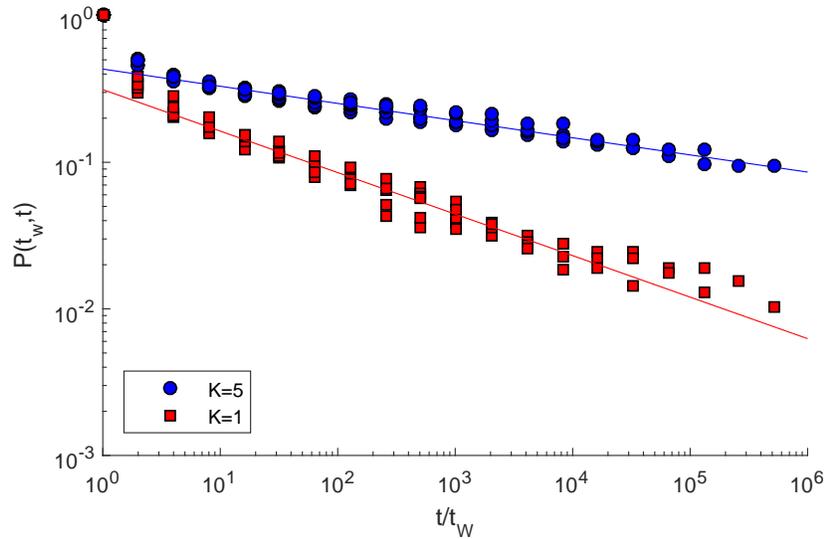}
	\caption{(Color on line) same data as in figure~\ref{fig:survival-rate}, now plotted as a function of $t/t_w$.
	$K=1$ data
	are shown by squares, and $K=5$ data by circles.
	 The lines are least squares fits to power-laws $y = a(t/t_w)^b$. 
	 where $b = -0.283(14)$ and $b = -0.117(6)$  for $K=1$ and $K=5$, respectively.
	 All available cohorts have been used in the fits.}
	\label{fig:t-on-tw}
\end{figure}
Since pure or $t/t_w$  scaling  is known to hold approximately in many aging systems of physical
nature, see Ref.~\cite{Sibani10} and references therein, it
is interesting to investigate its applicability to a model of biological evolution such as the TNM.
Figure~\ref{fig:t-on-tw} shows that plotting our  persistence data  as a function of $t/t_w$
produces  a good data collapse.  The lines are fits to a power-law $y(t/t_w) = a (t/t_w)^b$.
For $K=1$ the exponent is $b = -0.283(14)$ and for $K=5$ it is $b = -0.117(6)$.z
Three comments are in order: first, independently of the degree of inheritance, Eq.~\eqref{survival} shows that 
the life-time distribution lacks a finite average. Second, we see that species created at a late
stage of the evolution process (large $t_w$) are more  resilient than those created early on,
implying that the rate of quakes decreases in time.
Third, the exponent of the persistence decay is more than halved 
when $K$ goes from $1$ to $5$, 
clearly showing that 
  inheritance  produces a more robust ecology where  species  live longer.

\section{Conclusion and outlook}
In the TNM version introduced in this work,  a mutant  inherits part of 
the interactions of its parent, with the amount of modification controlled by an
integer  parameter $K$.
Independently of $K$,  the genome remains a point in an $L$ dimensional  hypercube,
and the value of $K$ has only  insignificant  effects 
of  the distribution of the non-zero couplings linking different species.
Finally, our approach does not require the storage and 
manipulation of huge sparse matrices.
We concur with Refs.~\cite{Sevim05, Laird06} that  trait inheritance does not radically changes the 
basic properties of the TNM:  irrespective of the value of  $K$, we see a decelerating  aging dynamics
where sudden quakes lead to considerable  rearrangements of the network structure of the TNM ecology.
More specifically,  independently of $K$ the  aging dynamics of the TNM    is
characterized by persistence curves which decay as powers of  the ratio
$t/t_w$,  a  scaling form
known as \emph{pure aging}  in complex systems of  physical origin~\cite{Sibani14}.
Increasing  the degree of inheritance has nevertheless   both structural and dynamical effects:
 The sub-net of core
species becomes larger and more highly populated, the species abundance distribution is better
approximated by the log-normal distribution found in  many experimental data,  and 
the decay of the species survival probability becomes noticeably slower.
 
Trait inheritance generates TNM
core species which are related to each other.
In real ecologies, be they of  biological or socio-economic nature,
families of related extant species  are the norm rather then the exception,
and the selective pressure generated a varying external parameter
 is more  likely to change the relative population of these related species than to destroy the 
ecosystem itself.
We expect  the added flexibility to external variations to be a property of the TNM
with trait inheritance which is worth of further investigations.

\begin{acknowledgments}
PS would like to thank Peter Salamon for motivating part of this work
and  Rudy Arthur for several stimulating  discussions on  TNM dynamics.
\end{acknowledgments}

 \end{document}